\documentclass[lettersize,journal]{IEEEtran}
\usepackage{amsmath,amsfonts}
\usepackage{algorithmic}
\usepackage{algorithm}
\usepackage{array}
\usepackage{textcomp}
\usepackage{stfloats}
\usepackage{url}
\usepackage{verbatim}
\usepackage{graphicx}
\usepackage{cite}
\usepackage{caption}
\usepackage{subcaption}
\usepackage{xcolor}
\usepackage{multirow}
\usepackage[normalem]{ulem}
\newcommand{\tabincell}[2]{\begin{tabular}{@{}#1@{}}#2\end{tabular}}

\begin{document}

\title{Massive MIMO Channel Measurement Data Set for Localization and Communication}

\author{Achiel Colpaert, Sibren De Bast, Andrea P. Guevara, Zhuangzhuang Cui, and Sofie Pollin
 \thanks{All authors are with WaveCoRE in Department of Electrical Engineering (ESAT), Katholieke Universiteit Leuven (KU Leuven), Leuven, Belgium.}}

\maketitle

\begin{abstract}
Channel state information (CSI) needs to be estimated for reliable and efficient communication, however, location information is hidden inside and can be further exploited. This article presents a detailed description of a Massive Multi-Input Multi-Output (MaMIMO) testbed and provides a set of experimental location-labelled CSI data. In this article, we focus on
the design of the hardware and software of a MaMIMO testbed 
for gathering multiple CSI data sets. We also show this data can be used for 
learning-based localization and enhanced communication research. The data set presented in this work is made fully
available to the research community.
We show a CSI-based joint communication and sensing processing pipeline can be evaluated and designed based on the collected data set. 
Specifically, the localization output obtained by a convolutional neural network (CNN) trained on the data sets is used to schedule users for improving the spectral efficiency (SE) of the communication system. Finally, we pose promising directions on further exploiting this data set and creating future data sets. 
\end{abstract}


\section{Introduction}
MaMIMO is an established technology used in the fifth-generation (5G) communication networks to improve reliability and SE by means of space-time diversity \cite{Larsson14}. In MaMIMO communication, the channel state information (CSI) between a large number of antennas in the base station (BS) and serving users is accurately estimated for effective transmission. MaMIMO systems are able to concentrate the signal power on the user and they enable efficient use of the spectrum. Therefore, it is useful to analyze this MaMIMO channel information to further understand and improve our communication systems. This channel information can be gathered by either conducting simulations of an environment or performing real-life measurements. Existing channel simulation techniques mainly rely on stochastic geometry-based channel models or ray-tracing techniques. These models allow for the generation of a large number of channel samples at any location or configuration and are often used when adopting data-driven approaches in research like in \cite{Wen18, Widmaier19}. However, stochastic channel models do not use the exact geometric properties for a given scenario, they merely take the relative position of the antenna elements into account, and all other parts of the environment are based on random clusters consisting of scatterers interacting with the wireless channel. As a result, this method cannot be used to evaluate performance in a real-life environment, which is especially important when considering joint communication and sensing. In addition, the virtual environment in the ray-tracing simulation must be reconstructed extremely accurately as different objects and materials will interact with propagating waves. Moreover, a ray-tracer has a highly computational cost, resulting in a low efficiency in generating a large number of data samples. Therefore, this article aims to employ a more practical method to accumulate an accurate location-labelled CSI data set, i.e, channel measurements, considering the high accuracy and low complexity. Previously, researchers have measured and published MaMIMO channel data sets \cite{ba_mimo, Decurnige18, Widmaier19}, 
however these often contain a low number of sample points or they are virtual data sets where channel measurements are performed sequentially and later combined, and these methods fails to encapsulate the actual synchronized channel state at different antennas. Therefore, to measure the actual synchronized channel between MaMIMO BS and its users, a MaMIMO testbed is required with real-time channel measurement capabilities over space, frequency and time. Furthermore, collecting a large amount of spatially labelled CSI data samples also poses several other challenges. 

One challenge is that gathering the CSI for mobile users is a time-consuming task since it requires moving them manually. For each sample in the data set, the location and CSI for a mobile user need to be captured and recorded simultaneously \cite{zhang_csi}. Another challenge is the accuracy of spatial labels. Measuring the location of the users manually 
generally only goes up to cm precision \cite{Widmaier19}, can be prone to human error and is a slow process. 
For outdoor location, Global Navigation Satellite System (GNSS) can be used for labelling, leading to an average accuracy of $3m$ when using a normal accuracy GNSS system. An improved system was used to gather the MaMIMO data set in \cite{Decurnige18} resulting in cm precision. However, this requires expensive hardware and software and is only available outdoors.
An indoor solution is to utilize an indoor positioning system for millimetre precision.
Then, additional design is needed for supporting automatic measurement by integrating a proprietary positioning system that can reach mm-level accuracy. In general, our main contributions can be summarised as follows. \textit{First}, for data set generation, we develop a measurement system, using the MaMIMO testbed and a positioning system, which can collect a massive data set automatically and accurately. \textit{Second}, 
we use the data set for two practical localization and communication solution research problems:
one is to use the CSI obtained in the communication process to localize users, the other is to enhance the communication performance by user scheduling based on localization information. 
\textit{Third}, for data set exposure, the data set is fully open to the research community and can be accessed from the IEEE DataPort platform \cite{sib_data}. The data set is tailored for 
further use with data-driven approaches such as Neural Networks and has already been used by other researchers such as in \cite{Guo2022, Ranjbar2022}.

The remainder of the article is as follows. The measurement MaMIMO testbed is introduced in Section \ref{sec:testbed}. Afterwards, the developed extensions of the testbed are presented in Section \ref{sec:extension}. In Section \ref{sec:dataset}, a detailed measurement campaign for data set collection is presented. Section \ref{sec:applications} first visualises the collected data sets with the received power after Multiple Input Single Output (MISO) processing for different array topologies.
Then, by employing a CNN, localization results are presented, and further used for improving the user selection and SE of communication systems with simple Maximum Ratio Transmission (MRT) precoding. Section \ref{sec:future} and Section \ref{sec:dataset_conclusion} poses future directions and concludes the article, respectively. 

\section{
Massive MIMO Testbed Design}
\label{sec:testbed}

In this section, an overview of the distributed KU Leuven MaMIMO testbed is presented, including the hardware and software. 
  
\subsection{The Hardware}

The testbed hardware mainly consists of Software Defined Radios (SDR) from National Instruments (NI), the NI USRP-2942R.
In this article, they are referenced as universal software radio peripherals (USRPs). We first present the BS composition, followed by the User Equipment (UE) description.

\subsubsection{The Base Station}

The BS is equipped with 64 antennas, where we use 32 USRPs as Remote Radio Heads (RRHs), each with two RF chains. The USRPs are controlled by a central processing unit (CPU) located in the main chassis of the BS. 
All devices are mounted in two server racks, combining the devices into one testbed. Fig.~\ref{fig:setup} shows a picture of the assembled BS at the back. The KU Leuven MaMIMO testbed supports the flexible antenna deployment, such as uniform linear array (ULA), uniform rectangular array (URA), and distributed array (DA).

\subsubsection{The User Equipment}

The UE is also based on the NI USRP-2942R. Since one USRP provides two RF chains, one USRP can be used for two users.
The UEs are connected to a host PC using a PXIe cable for a digital link to control and forward the transmitted and received data. The UEs can be connected to the BS using coax cables 
in this way, the UEs can achieve a perfect synchronisation in time and frequency with the BS which is a prerequisite for measurements. However, the provided UEs are static and bulky, and a automated positioning and mobility system is still required.

\subsection{The Software: LabView MIMO Application Framework}

All of the presented hardware is driven using the provided NI MIMO application framework. This is a LabView software project that implements all functionality to communicate between the BS and UEs. Researchers can implement alterations or extensions directly in the provided project. 

The framework uses a LTE-based frame structure using Time Division Duplexing (TDD) and numerology 0, which represents the standard parameters for the first configuration of waveforms, each subcarrier has $15kHz$, and the symbol duration is $7.1\mu s$. The packets are modulated using orthogonal frequency-division multiplexing. UL, and DL pilots are sent in different time-slots.

The UL pilot is used to estimate the channel at the BS side and in total 12 different UL pilot symbols are available in the framework, one for every user. 
Every UL pilot uses 100 subcarriers of the 1200 available subcarriers. The 12 different UL pilots are frequency interleaved, each pilot only uses one of the 12 subcarriers per resource block. In this way, all 12 users can send a pilot during the same time slot. A channel is estimated for all 64 antennas and for 100 different subcarriers. Furthermore, In-phase and Quadrature components are measured. Therefore, one channel sample of one user is represented as $64 \times 100$ matrix of complex values.

A shortcoming of the framework is that no channel logging functionality is implemented. One solution to log the CSI of the users is to intercept the channel estimation as it is sent from the MIMO processors to the CPU. However, at the CPU the channel is only updated every $100ms$, as the CSI at this location is only used to visualize the channel for debugging purposes. A better way is to log the CSI on the FPGA where all samples are processed, resulting in 2000 updates per second. However, this requires the development of code that has to run on the FPGAs, making it time-consuming to implement. For the purposes of the channel measurements delineated in this work, the channel capturing at the CPU side is used as this is sufficient. Example study using the same facility but with denser data logging is for instance Sakhnini \textit{et al.} \cite{Sakhnini2022} who integrates radar sensing in a MaMIMO system.

\begin{figure*}
\centering
\includegraphics[width=0.85\textwidth]{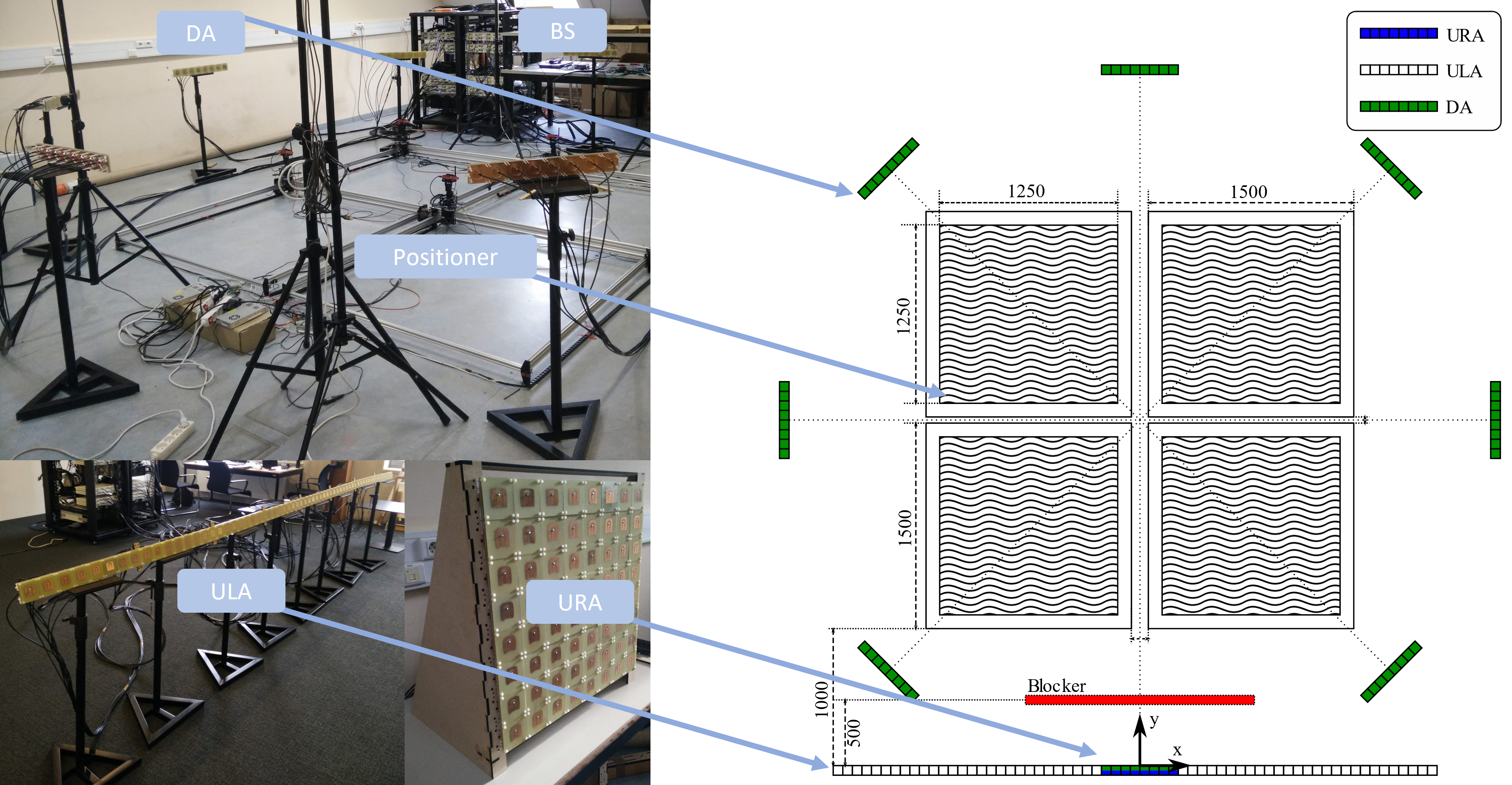}
\caption{
Overview of the setup. On the left pictures of the hardware including the BS, its antenna configurations: ULA, URA, DA; and the positioning system.
On the right, a top view schematic layout of the measurement configurations. The arrows indicate the matching hardware elements from the left. Three antenna topologies are created. For the URA scenario, the eight-by-eight antenna array (blue) is located in front of the ROI. Using the URA, both a LoS and nLoS data set are created. A metal blocker was placed in front of the array to create the nLoS scenario. For the ULA configuration, a linear array is deployed in front of the ROI. For the DA scenario, eight arrays of eight antennas are distributed in an octagonal shape around the users. All measures are in millimetres.}
\label{fig:setup}
\end{figure*}

\section{Testbed Implementation for Automated Measurements}
\label{sec:extension}

We describe the developed extension to automate the testbed implementation, enabling it to perform massive and accurately-labelled CSI measurements.

\subsection{CNC-enabled Automatic User Placement}

In order to perform automated measurements with the testbed, the mobility pattern of the UEs has to be designed. 
Thus, we connected long coax cables to the antenna ports of the UEs' USRPs and 
mounted only the antenna on a robotic system. In this way, only the antenna has to move around to change the position of the user, instead of moving the full device. 

To move the antennas over a given area, a suitable robotic systems has to be selected. As the goal of the data set is to develop ultra accurate positioning methods, the robotic positioners have to be able to position the antennas with a very high level of accuracy. Furthermore, as the second target is to create a data set with a very high amount of data points, the positioners should be very reliable that the locations are still correct after a long time of operation. To this means, Computer Numerical Controlled (CNC) tables were selected. They are designed to achieve a very high positioning accuracy and reliability. 

The selected CNC-tables are the OpenBuilds ACRO 1515. This positioner reliably covers a $1250mm \times 1250mm$ area.. Furthermore, the reported error of the positioner is less than $0.1mm$. In total four of these positioners are used, shown in Fig.~\ref{fig:setup}. Custom brackets hold the legs of multiple positioners so that their relative position is known. Using the positioners, the location of a user can be altered without manual intervention. The positioners make it possible to scan complete areas gathering the CSI of the users.  

\subsection{Automated Channel Capture Triggers using TCP}

With the previously presented testbed extensions, the position and orientation of the users can be altered automatically. The next step is to automate the process of capturing a CSI sample at the BS. For this purpose, the BS software is extended to receive Transmission Control Protocol (TCP) packets to automatically trigger the capture of a CSI sample. Each TCP packet contains six bytes of data, which are used as filename to save the captured CSI sample. In this way, the BS can be triggered from a remote device to perform an automated measurement and label the sample accordingly. The developed code intercepts the channel estimate and writes the CSI sample to a binary file. The complete set of binary files make up the measured data set.

\section{Open Data Set}
\label{sec:dataset}

The extended testbed was used to capture two data sets using three different antenna topologies, which are presented first. The first data set is called the dense data set (DDS). This data set is created by scanning the full range of the xy-positioners using a very dense grid-pattern, both in a Line-of-Sight (LoS) and non-Line-of-Sight (nLoS) scenario. The second data set is the nomadic data set, which focuses on the influence of moving objects in the environment.

To support transparency in research and improve repeatability of research, these two data sets are published under an open-access license and can be freely downloaded through the IEEE DataPort platform \cite{sib_data}. As a result, the data can be used by other researchers to develop or validate new signal processing algorithms using measured data such as by Ranjbar \textit{et al.} \cite{Ranjbar2022}, who explore Cell-free ORAN architectures using our data set. This is a major contribution to the research community as MaMIMO testbeds are not common and a freely available MaMIMO CSI data set was missing. For the remainder of this work we focus on the details and results of the LoS case of the DDS. More details and studies with the other data sets can be found in \cite{DeBast2022}.

\begin{figure*}
    \centering
    \includegraphics[width=0.75\textwidth]{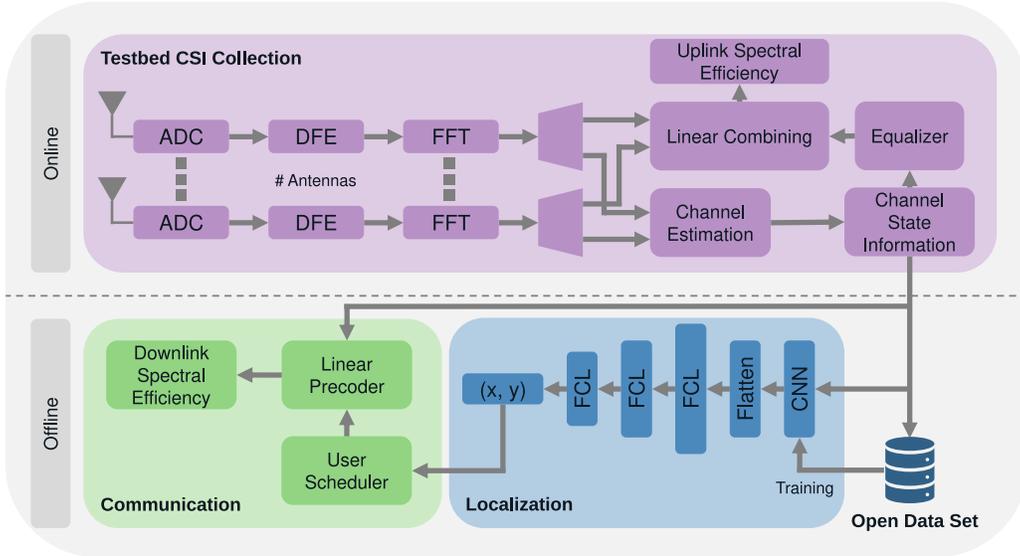}
    \caption{Overview of the complete joint communication and sensing system. The top side of the figure shows how the CSI data is gathered and stored in the Open Data Sets described in this work. The localization section in blue shows the CCN structure used for extracting xy-locations from the CSI data. The communication section in green describes how CSI and xy-location are used to schedule and precode data for different users. The output SE is used to evaluate the system.} 
    \label{fig:arch}
\end{figure*}
\subsection{Overview of the Scenario}

The data sets were recorded in an indoor location, the MaMIMO lab at ESAT, KU Leuven. The xy-positioners were placed in a rectangle in the middle of the room. The space covered by the positioners is called the Region-of-Interest (ROI). For both data sets, three distinctive antenna topologies are deployed. The three antenna topologies and their locations can be seen in Fig. \ref{fig:setup}. The three antenna deployments are:

\begin{itemize}
  \item \textbf{Uniform Rectangular Array (URA):} This array of eight by eight patch antennas is deployed in front of the ROI. The array is placed with the middle of the array at $1m$ height above the ground. The distance between the antenna elements centre is $7cm$. The array directly faces the ROI, all possible user locations are in LoS of all antenna elements.
  \item \textbf{Uniform Linear Array (ULA):} This array of 64 antennas is placed in front of and facing the ROI. The height of patch antennas' centre is set to $1m$. With a distance of $7cm$ between the centres, the resulting array has a length of $4.48m$. The array is fully in LoS for all user positions.
  \item \textbf{Distributed Array (DA):} The 64 antenna elements are distributed over eight ULAs of eight antennas. The arrays are distributed around the ROI in an octagonal shape around the ROI. All antenna elements are placed on the same height of $1m$ and are facing the middle of the ROI.
\end{itemize}

The right side of Fig. \ref{fig:setup} shows a top view schematic layout of the lab with the different antenna topologies. The BS's antennas are indicated by the small squares, with each colour depicting another scenario. The waved area inside the four squares is the ROI in which the antennas can be moved. For the labelling of the positions, the middle of the URA was chosen as the origin of the local coordinate system. This location is shown by the xy-axis on the figure. All user positions in the data set are measured in millimetres, referenced to this origin and axes. The exact measurements of the scenarios are shown on the figure, while the exact coordinates of the users and antennas are provided as a list accompanying the published data sets.

For all scenarios, the same configuration of the BS was used. The frequency used is $2.61GHz$ with a transmit power of $18.5dBm$ at the user and a receive gain of $15dBm$ at the BS. To ensure a perfect synchronization between the users and the BS, the synchronization signals were transferred using coax cables.

\subsection{Dense Data Set}

The goal of this data set is to create a very large labelled data set containing as many CSI samples of different locations as practically possible. Previous data sets \cite{Widmaier19} collected to train localization algorithms contained up to a few thousand samples. Therefore, we want to maximize the number of locations in the data set. The grid over which the antennas move has a resolution of $5mm$. This way, for each positioner, a 251-by-251 grid was scanned, resulting in 63,001 samples per positioner, or 252,004 sampled positions total per antenna topology. At every node in the grid, the positioner stopped for $0.5s$ in order for the BS to record the channel of the users being static. 

Moving the antennas to the next node in the grid and capturing a CSI sample takes for this setup on average $0.7s$, therefore, to complete the full measurement, around 12.5 hours are required. This duration sets the limit of the size of the measurement. Long measurements require a very high reliability of the measurement setup. The xy-positioners move cables around, decreasing the reliability of the setup as the cables get tangled. As a result, in order to complete a full dense measurement as described here, multiple tries were required. 

\section{Localization and Communication Analysis}
\label{sec:applications}

In this section, we discuss several applications of the DDS. First, to verify the correctness of the data set, the data set is visualized in different ways. This will give a visual indication on the correctness of the data set \cite{guevara21}. Next, a CNN is designed and utilized to extract location information from the CSI data \cite{DeBast20}. Finally, this location information is used to develop a location based user scheduling algorithms, resulting in all the building blocks required for a joint communication and sensing system\cite{DeBast2022}. An overview of the complete system can be seen in Fig. \ref{fig:arch}.

\subsection{Visualising MaMIMO precoders}

Using the DDS, we can visualise the beamforming patterns and visually verify the correctness of the data set. As precoding method, MRT is chosen. This beamforming technique is well understood as it maximises the array gain. When applying MRT beamforming, we expect users with a high channel correlation to the channel of the targeted user to receive high power.

\begin{figure*}
    \begin{subfigure}{0.33\textwidth}
        \centering
        \includegraphics[width=.95\textwidth]{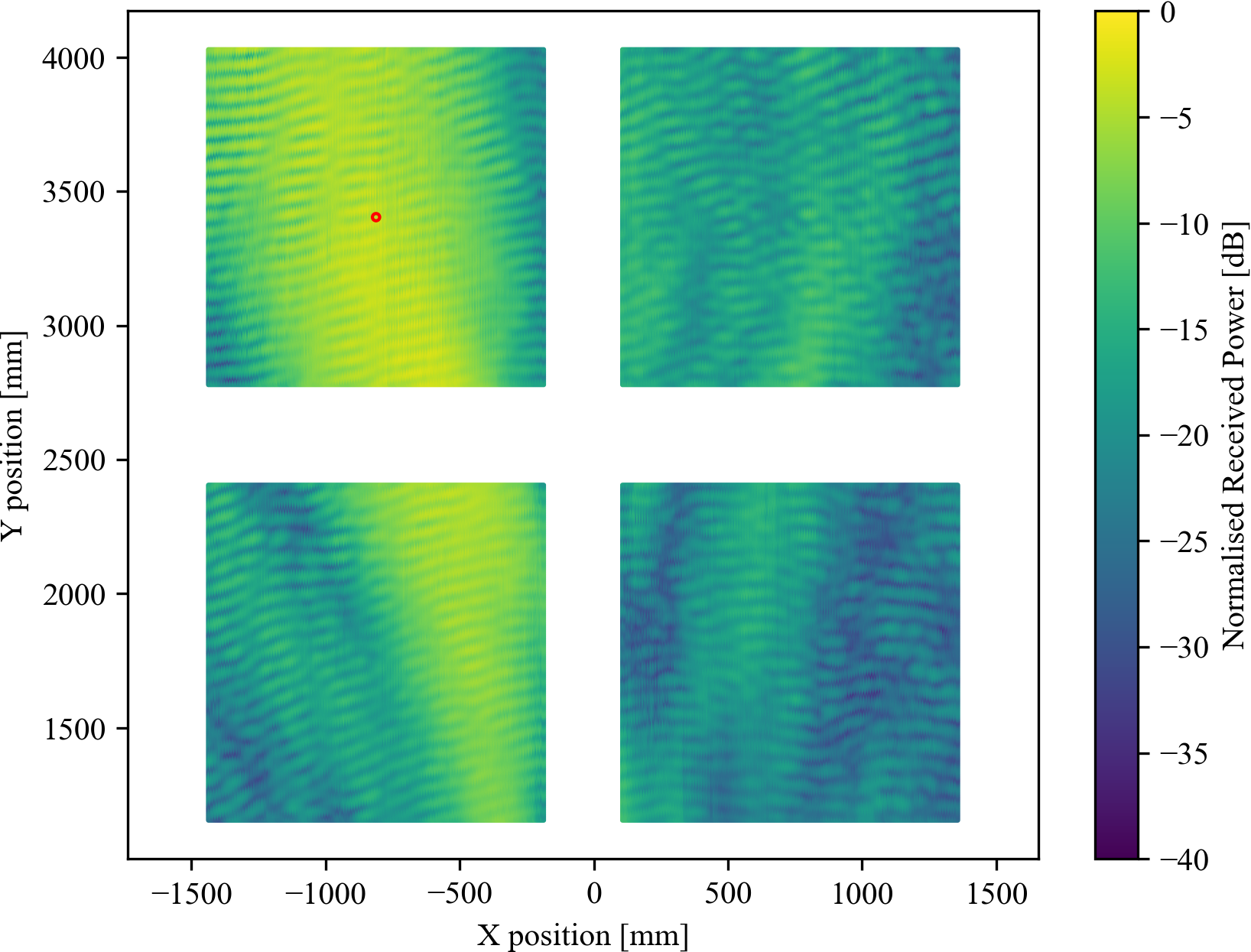}
        \caption{(a) URA}
        \label{fig:MR_URA}
    \end{subfigure}
  \begin{subfigure}{0.33\textwidth}\centering
  \includegraphics[width=.95\textwidth]{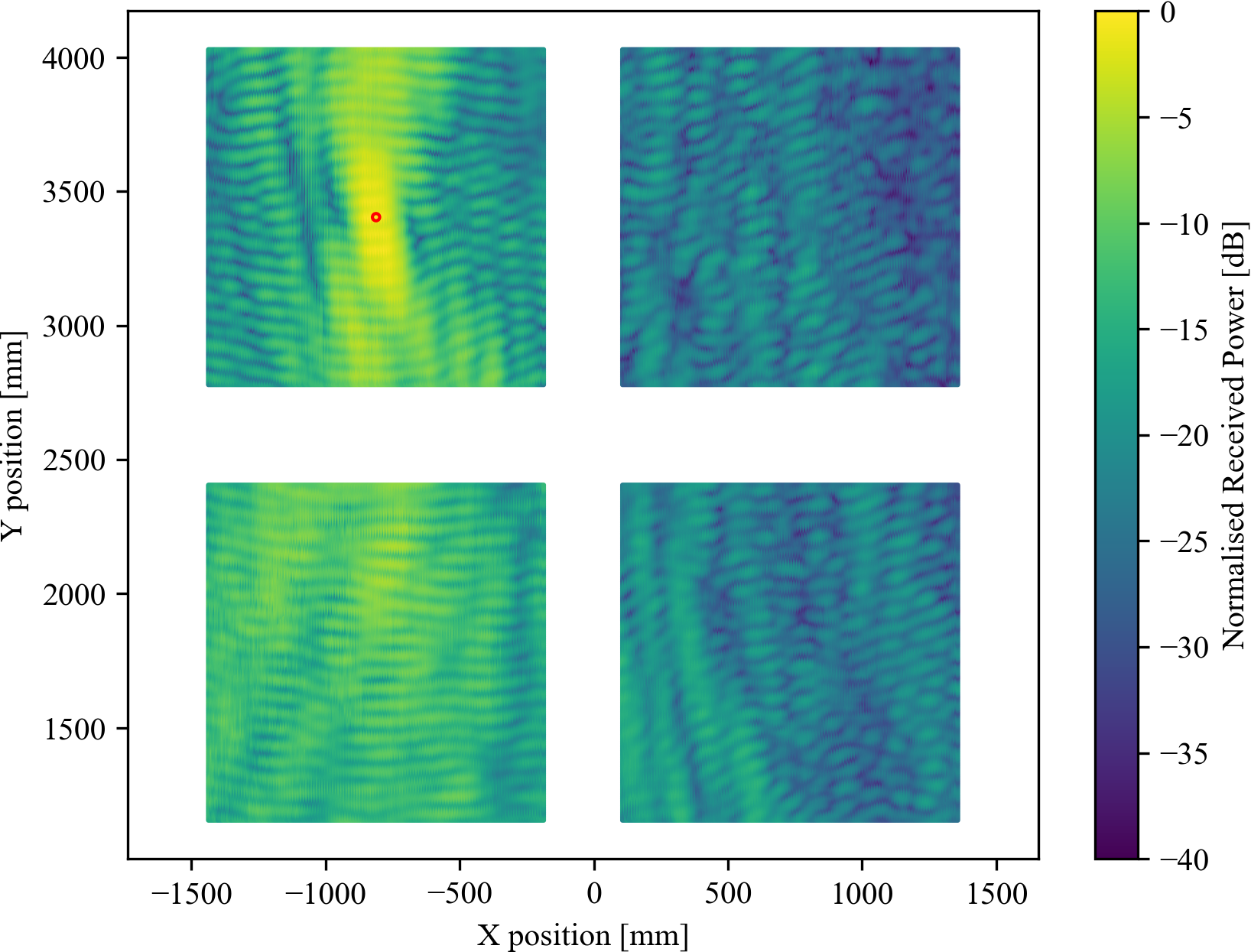}
  \caption{(b) ULA}
    \label{fig:MR_ULA}
  \end{subfigure}
  \begin{subfigure}{0.33\textwidth}\centering
  \includegraphics[width=.95\textwidth]{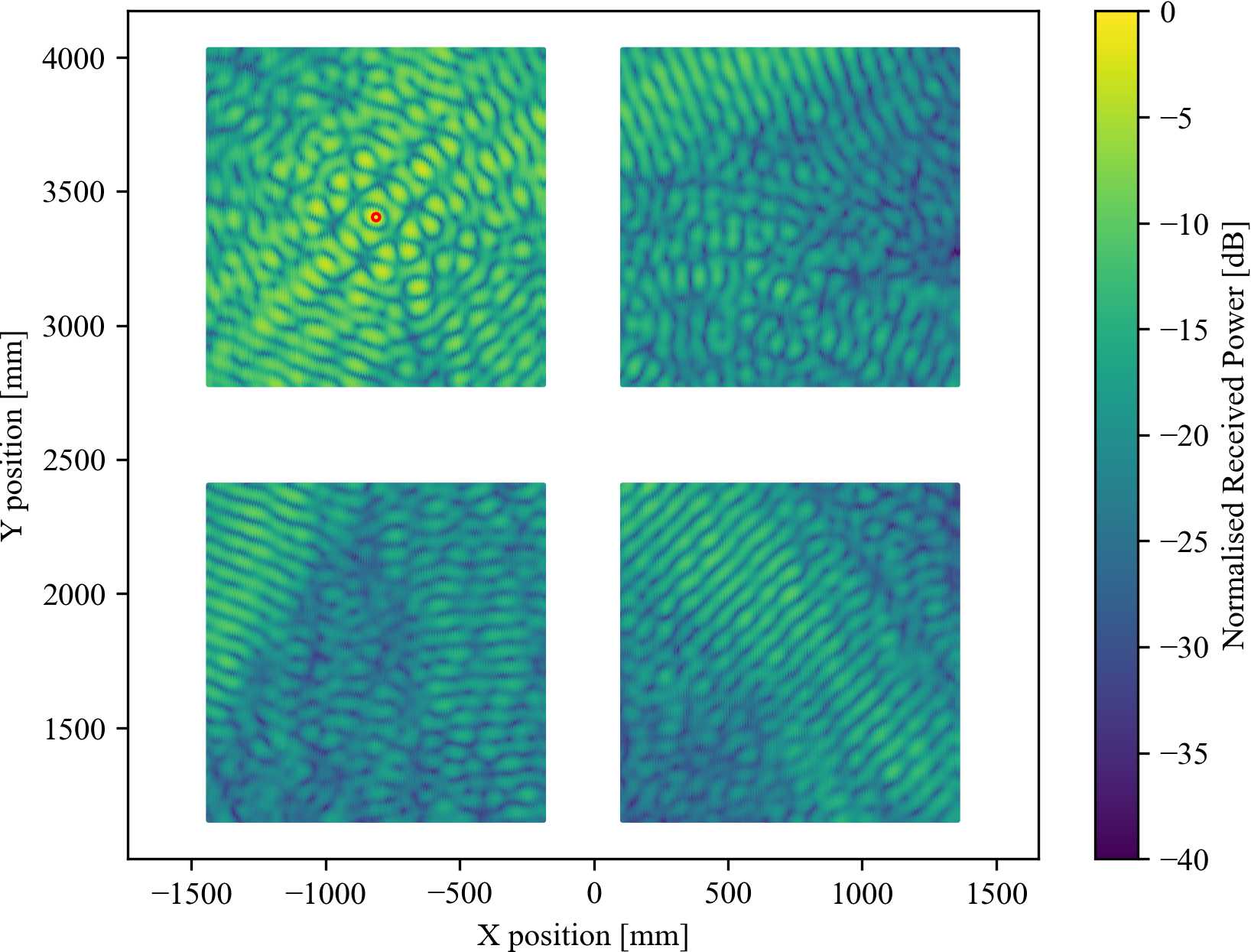}
  \caption{(c) DA}
  \label{fig:MR_DIS}
  \end{subfigure}
    \caption{Normalized received signal strength by applying MRT beamforming towards the user indicated by the red dot 
    \cite{guevara21}.  }
    \label{fig:MR_beamforming}
\end{figure*}

Fig. \ref{fig:MR_beamforming} show the normalised received power for all 252,004 samples of each measurement in the DDS when employing MRT beamforming towards one user. The location of the targeted user is indicated by the red dot. Power in these figures is normalised by dividing all powers by the maximum power measured in the area over all four antenna topologies, so that the different topologies can be compared. These figures are reprinted from \cite{guevara21}.

The images show a consistent power distribution over the scanned area, without any discontinuities, hinting that the measurements are indeed meaningful and correct. Moreover, the fine grained resolution of the measurements enables the visualisation of the small-scale fading effects, something that is often not included in stochastic channel models.

In \cite{Guevara19}, we used the DDS to compare the performance of the communication system for the three antenna configurations. The metric is the number of users that can be served simultaneously when using Zero-Forcing precoding. The results show that for this scenario the DA antenna configuration could serve simultaneously 18 more users than the URA, due to the intrinsic lower channel correlation obtained by distributing the antennas. See TABLE \ref{tab:results}.

\subsection{CNN-based positioning}
CNNs are a type of Neural Network that apply convolutions on the data with trainable filters to extract features and information. In this case, the goal is to accurately position users of a MaMIMO communication system, using the CSI. The CNN is used to extract features from the CSI and convert it to a low dimensional vector, which is used to determine the position of the associated user. The spatial features are complex, as the CSI contains information about both direct and indirect propagation paths. Therefore, a deep CNN is designed to achieve this goal. Figure \ref{fig:arch} shows the architecture of the proposed CNN. It consists of two main parts, a CNN part followed by a Fully Connected Neural Network. The results presented in TABLE \ref{tab:results} show a high position performance with cm-accuracy. Details can be found in \cite{DeBast20}.

\subsection{{Localization-Enhanced Spectral Efficiency}}
Using the location information of the users a communication system can improve its SE in several ways, for example, location based user allocation or location based user scheduling. These techniques can improve individual SE because users will interfere less with each other. In our work we explore location based user scheduling using the DDS. We propose a novel user scheduling technique called Distance Enhanced Flocking (DEF) that relies on the user location. 

Our results are compared to the popular user selection algorithm called Semi-orthogonal User Selection (SUS), which selects users based on the orthogonality of their channel to each other \cite{Yoo06}. SUS operates well when the number of users is large but may be sub-optimal when the pool of users is small. Our proposed technique, DEF, is designed to accomplish two objectives for smaller pools of users. The first objective is to ensure that users that are located closely together are not clustered in the same group. The second is to keep a low computational complexity as these algorithms must be able to run fast and potentially divide a large number of users into groups. The main idea of DEF is to sort the users based on a minimal distance in between two sequential users. The results confirm that the distribution of antennas help to increase the SE of the system. Interestingly, for this type of user allocation the SE results show an small variation between ULA and DA, see TABLE \ref{tab:results}. More details on the algorithm can be found in \cite{DeBast2022}.

\begin{table}[!t]
    \centering
     \caption{CSI Data Set for Localization and Communication Analytical Results. The reported SE is obtained through user scheduling using the SUS algorithm \cite{Yoo06} and our proposed DEF algorithm.}
    \begin{tabular}{|l|c|c|c|c|}\hline
  & Localization & \multicolumn{3}{c|} {Communication} \\ \cline{2-5}
  Topology  & \multirow{2}{*}{\tabincell{c}{Mean Error\\ (mm) \cite{DeBast20}}}  & \multirow{2}{*}{\tabincell{c}{ Max. Served \\Users\cite{Guevara19}}}  & \multicolumn{2}{c|} {\tabincell{c}{SE \\(bits/s/Hz)}}\\ \cline{4-5}
     & & & SUS \cite{Yoo06}  & Our DEF \\ \hline
    URA & 20.73 & 25 & 20.85 & 22.04\\ \hline
    ULA & 21.11 & 38 & 24.14 & 24.41\\ \hline
    DA  & 18.04 & 43 & 23.87 & 24.12\\ \hline
    \end{tabular}
    \label{tab:results}
\end{table}

\section{Future Directions}
\label{sec:future} 
In this section, we identify three research directions for further exploitation, validation, and application of the data set.

\subsection{Applying Deep Learning on CSI Time Series}
Our recent extensions of the testbed are capable of recording up to 2000 CSI samples per second for multiple users. This can be used to record refined time series of MaMIMO CSI samples. When the location of a user is time dependent, i.e. there is a very large dependency between the previous position and the current position, the time-varying features between multiple CSI samples can be utilized to further improve the localization and tracking performance. To this end, Recurrent Neural Networks (RNN) and 
transfer learning can be used. 
These advanced methods can potentially unlock practical mm-level localization.

\subsection{Extensive Data Sets for Larger and More Complex Areas}
Since the current work is only validated using an indoor data set spanning a relatively small area, we can improve on our current data sets. First of all, a larger area can be considered, with locations at different heights, enabling 3D localization. Next, a data set in an outdoor scenario can be recorded, optionally including measurement at different heights by mounting the user on drone. When recording a larger data set, more complex propagation conditions can be considered, where both LoS and nLoS locations are included in the data set. In our latest work, a mobile user compatible with the testbed was developed using a compact embedded SDR, which enables large-area mobile measurement campaigns, where the user is not bounded by the synchronization cable between UE and BS. Extended data sets can also fuel important domain adaptation research, which is a key bottleneck when considering Deep Neural Network-driven physical layer design.

\subsection{Enabling Data Set for Joint Communication and Sensing}
The current bulk of work has focused separately on communication and localization. However, with longer and faster data measurements it becomes feasible to also consider the Doppler domain, and study active and passive sensing. By using the presented DDS and our proposed architecture, more flexible and advanced joint sensing and communication systems can be created virtually using real data, allowing for a thorough analysis of any novel trials with realistic environments and targets.  

\section{Conclusion}
\label{sec:dataset_conclusion}

In this article, we presented a MaMIMO testbed design and implementation, which are the measurement foundation for creating massive CSI data sets. The flexible antenna deployment capabilities of the KULeuven MaMIMO testbed allow for distinct antenna array topologies. Moreover, for an automatic and accurate measurement campaign, we developed a central controller that accurately moves four users through the testing scenario while directing the BS to conduct channel measurements at the different locations, which prompts us to obtain the DDS with 252,004 location-labelled CSI samples for a grid of $5mm$ in space.

For real applications, the DDS was used to generate the normalized received signal strength over the scanned area when the beamforming towards a specific location. The results show no discontinuities and correspond with the theoretical understanding of the used MRT precoder, demonstrating all recorded CSI samples are accurate. A CNN was trained using the data set to extract xy-location information out of CSI data. Then, this location knowledge was used to develop a location-based user scheduling technique. The bulk of work based around the presented data sets results in complete building blocks for joint communication and sensing systems. Finally, future directions for this open data set and new potential open data sets are presented. This work shows the value of such open data sets for the scientific community.

\section{Acknowledgements}
This work was funded by European Union's Horizon 2020 under grant agreement no. 101017171 (Machine Learning-Based, Networking and Computing Infrastructure Resource Management of 5G and Beyond Intelligent Networks - MARSAL project), no.  101013425 (REsilient INteractive applications through hyper Diversity in Energy Efficient RadioWeaves technology - REINDEER project). The software for the positioners was developed by Jona Beysens.

\bibliography{papers}

\begin{thebibliography}{10}
\providecommand{\url}[1]{#1}
\csname url@samestyle\endcsname
\providecommand{\newblock}{\relax}
\providecommand{\bibinfo}[2]{#2}
\providecommand{\BIBentrySTDinterwordspacing}{\spaceskip=0pt\relax}
\providecommand{\BIBentryALTinterwordstretchfactor}{4}
\providecommand{\BIBentryALTinterwordspacing}{\spaceskip=\fontdimen2\font plus
\BIBentryALTinterwordstretchfactor\fontdimen3\font minus
  \fontdimen4\font\relax}
\providecommand{\BIBforeignlanguage}[2]{{%
\expandafter\ifx\csname l@#1\endcsname\relax
\typeout{** WARNING: IEEEtran.bst: No hyphenation pattern has been}%
\typeout{** loaded for the language `#1'. Using the pattern for}%
\typeout{** the default language instead.}%
\else
\language=\csname l@#1\endcsname
\fi
#2}}
\providecommand{\BIBdecl}{\relax}
\BIBdecl

\bibitem{Larsson14}
E.~G. Larsson, O.~Edfors, F.~Tufvesson, and T.~L. Marzetta, ``Massive mimo for
  next generation wireless systems,'' \emph{IEEE Communications Magazine},
  vol.~52, no.~2, pp. 186--195, 2014.

\bibitem{Wen18}
C.-K. Wen, W.-T. Shih, and S.~Jin, ``Deep learning for massive mimo csi
  feedback,'' \emph{IEEE Wireless Communications Letters}, vol.~7, no.~5, pp.
  748--751, 2018.

\bibitem{Widmaier19}
M.~Widmaier, M.~Arnold, S.~Dorner, S.~Cammerer, and S.~ten Brink, ``Towards
  practical indoor positioning based on massive mimo systems,'' in \emph{2019
  IEEE 90th Vehicular Technology Conference (VTC2019-Fall)}, 2019, pp. 1--6.

\bibitem{ba_mimo}
B.~Ai, K.~Guan, R.~He, J.~Li, G.~Li, D.~He, Z.~Zhong, and K.~M.~S. Huq, ``On
  indoor millimeter wave massive mimo channels: Measurement and simulation,''
  \emph{IEEE Journal on Selected Areas in Communications}, vol.~35, no.~7, pp.
  1678--1690, 2017.

\bibitem{Decurnige18}
A.~Decurninge, L.~G. Ordóñez, P.~Ferrand, H.~Gaoning, L.~Bojie, Z.~Wei, and
  M.~Guillaud, ``Csi-based outdoor localization for massive mimo: Experiments
  with a learning approach,'' 2018.

\bibitem{zhang_csi}
Y.~Zhang, D.~Li, and Y.~Wang, ``An indoor passive positioning method using csi
  fingerprint based on adaboost,'' \emph{IEEE Sensors Journal}, vol.~19,
  no.~14, pp. 5792--5800, 2019.

\bibitem{sib_data}
\BIBentryALTinterwordspacing
S.~De~Bast and S.~Pollin, ``Ultra dense indoor mamimo csi dataset,'' IEEE
  Dataport, accessed Oct. 25, 2022. [Online]. Available:
  \url{https://dx.doi.org/10.21227/nr6k-8r78}
\BIBentrySTDinterwordspacing

\bibitem{Guo2022}
J.~Guo, C.-K. Wen, M.~Chen, and S.~Jin, ``Environment knowledge-aided massive
  mimo feedback codebook enhancement using artificial intelligence,''
  \emph{IEEE Transactions on Communications}, vol.~70, no.~7, pp. 4527--4542,
  2022.

\bibitem{Ranjbar2022}
V.~Ranjbar, A.~Girycki, M.~A. Rahman, S.~Pollin, M.~Moonen, and E.~Vinogradov,
  ``Cell-free mmimo support in the o-ran architecture: A phy layer perspective
  for 5g and beyond networks,'' \emph{IEEE Communications Standards Magazine},
  vol.~6, no.~1, pp. 28--34, 2022.

\bibitem{Sakhnini2022}
A.~Sakhnini, S.~De~Bast, M.~Guenach, A.~Bourdoux, H.~Sahli, and S.~Pollin,
  ``Near-field coherent radar sensing using a massive mimo communication
  testbed,'' \emph{IEEE Transactions on Wireless Communications}, vol.~21,
  no.~8, pp. 6256--6270, 2022.

\bibitem{DeBast2022}
S.~De~Bast and S.~Pollin, ``\BIBforeignlanguage{eng}{User localisation in
  massive mimo networks},'' Ph.D. dissertation, KU Leuven, 2022-10-14.

\bibitem{guevara21}
A.~P. Guevara, S.~De~Bast, and S.~Pollin, ``Weave and conquer: A
  measurement-based analysis of dense antenna deployments,'' in \emph{IEEE
  International Conference on Communications (ICC)}, 2021, pp. 1--6.

\bibitem{DeBast20}
S.~De~Bast, A.~P. Guevara, and S.~Pollin, ``Csi-based positioning in massive
  mimo systems using convolutional neural networks,'' in \emph{IEEE 91st
  Vehicular Technology Conference (VTC-Spring)}, 2020, pp. 1--5.

\bibitem{Guevara19}
A.~P. Guevara, S.~De~Bast, and S.~Pollin, ``Mamimo user grouping strategies:
  How much does it matter?'' in \emph{53rd Asilomar Conference on Signals,
  Systems, and Computers}, 2019, pp. 853--857.

\bibitem{Yoo06}
T.~Yoo and A.~Goldsmith, ``On the optimality of multiantenna broadcast
  scheduling using zero-forcing beamforming,'' \emph{IEEE Journal on Selected
  Areas in Communications}, vol.~24, no.~3, pp. 528--541, 2006.

\end{thebibliography}
\bibliographystyle{IEEEtran}




 



 \begin{IEEEbiographynophoto}{Achiel Colpaert} [IEEE Member] obtained his BSc and MSc degree in electrical engineering from KU Leuven, Belgium respectively in 2015 and 2017. Currently, he is working towards his PhD at KU Leuven, focusing on high-throughput wireless links for UAV applications. His main research interests are mmWave and outdoor wireless modelling focusing on measurement based work.
 \end{IEEEbiographynophoto}
 \begin{IEEEbiographynophoto}{Sibren De Bast} [IEEE Member] received the Ph.D. degree in electrical engineering from KU Leuven, Leuven, Belgium, in 2022. He is focusing on massive MIMO communications for aerial and ground robots, for which he created an experimental test and measurement environment. His current research interests include artificial intelligence for communications and localization.
 \end{IEEEbiographynophoto}
 \begin{IEEEbiographynophoto}{Andrea P. Guevara} [IEEE Member] obtained her BSc in Electronics and Telecommunications at the University of Cuenca, Ecuador in 2013. In 2015 she got her MSc in Telecommunications and the prize for the best academic performance at Research at King's College London, UK. In 2022 she got her Ph.D. degree at KU Leuven, her main interests are interference analysis and cooperation in MaMIMO systems.
 
 \end{IEEEbiographynophoto}

 \begin{IEEEbiographynophoto}{Zhuangzhuang Cui} [IEEE Member] received his Ph.D. degree from Beijing Jiaotong University (BJTU), Beijing, China, in April 2022. Currently, he is a postdoctoral research associate at KU Leuven, Belgium. From 2019 to 2021, he was a visiting scholar and Ph.D. student at several universities (UPM, NCSU, UCLouvain) in Spain, the USA, and Belgium. His research interests include channel modeling and non-terrestrial networks.
 \end{IEEEbiographynophoto}
 
 \begin{IEEEbiographynophoto}{Sofie Pollin} [IEEE Senior Member] received the Ph.D. degree (Hons.) from KU Leuven, in 2006. From 2006 to 2008, she continued her research on wireless communications at UC Berkeley. In 2008, she returned to IMEC to become a Principal Scientist at the Green Radio Team. She is currently a Full Professor with the Electrical Engineering Department, KU Leuven. Her research interests include networked systems that require ever more dense, heterogeneous, battery-powered, and spectrum constrained networks. She is a BAEF Fellow and a Marie Curie Fellow.
 \end{IEEEbiographynophoto}

\vfill

\end{document}